\documentclass[11pt]{article}
\pdfoutput=1
\usepackage{a4wide}
\usepackage{amsmath}
\usepackage{latexsym}
\usepackage{amssymb}
\usepackage{fancybox}
\usepackage{graphicx}
\usepackage{bbm}

\newtheorem{Thm}{Theorem}[section]
\newtheorem{Lem}[Thm]{Lemma}

\newenvironment{proof}{\noindent {\bf Proof:}}{$\Box$ \medskip}

\author{Deeparnab Chakrabarty\thanks{University of Waterloo (deepc@math.uwatrloo.ca)} \and Chinmay Karande\thanks{Georgia Institute of Technology (ckarande@cc.gatech.edu)} \and Ashish Sangwan\thanks{Georgia Institute of Technology (ashish.sangwan@gmail.com)}}
\title{The Effect of Malice on Social Optimum in Linear Load Balancing Games}
\date{}
\begin{document}

\maketitle
\begin{abstract}
In this note we consider the following problem to study the effect of malicious players
on the social optimum in load balancing games: Consider two players SOC and MAL controlling
$(1-\alpha)$ and $\alpha$ fraction of the flow in a load balancing game.
SOC tries to minimize the total cost faced by her players while MAL tries to maximize the same. 

If the latencies are linear, we show that this 2-player zero-sum game has a {\em pure} strategy Nash equilibrium. 
Moreover, we show that one of the optimal strategies for MAL is to play {\em selfishly}: let the $\alpha$ 
fraction of the flow be sent as when the flow was controlled by infinitesimal players playing selfishly 
and reaching a Nash equilibrium. This shows that a malicious player cannot
cause more harm in this game than a set of selfish agents. 

We also introduce the notion of Cost of Malice - the ratio of the 
cost faced by SOC at equilibrium to $(1-\alpha)\mbox{OPT}$, where OPT is the social optimum minimizing the cost
of all the players. In linear load balancing games we bound the cost of malice by $(1+\alpha/2)$.
\end{abstract}

\def\MAL{{\sc MAL }}
\def\SOC{{\sc SOC }}
\def\MALSOC{{\sc MALSOC}}
\def\SOCMAL{{\sc SOCMAL}}

\def\z{{\bf z}}
\def\y{{\bf y}}
\def\x{{\bf x}}

\def\a1{a_1}
\def\b1{b_1}
\def\x1{x_1}
\def\yos{y_1^*}
\def\ai{a_i}
\def\bi{b_i}
\def\xi{x_i}
\def\si{s_i}
\def\thi{\theta_i}
\def\yi{y_i}
\def\yis{y_i^*}
\def\s{\textbf{s}}
\def\x{\textbf{x}}
\def\y{\textbf{y}}
\def\sm{\mathbf{s}}
\def\yh{$\hat{\mathbf{y}}$}
\def\yhm{\hat{\mathbf{y}}}
\def\ys{$\mathbf{y}^*$}
\def\ysm{\mathbf{y}^*}
\def\soc{SOC}
\def\mal{MAL}
\def\socmal{SOC-MAL}
\def\malsoc{MAL-SOC}
\def\socmalm{\textrm{SOC-MAL}}
\def\malsocm{\textrm{MAL-SOC}}
\def\nash{$NASH$}
\def\nashm{C_N(1)}
\def\scale{SCALE}
\def\scalem{\textrm{SCALE}}
\def\opt{C_S(1)}
\def\optm{C_S(1)}
\def\sq{S_{Q}}
\def\sp{S_{P}}
\def\sps{S_{P}^*}
\def\spq{S_{Q}^*}
\def\ae{a_e}
\def\be{b_e}
\def\xe{x_e}
\def\ye{y_e}
\def\yes{y_e^*}
\def\bp{b_P}
\def\bq{b_Q}
\def\fp{f_P^*}
\def\fq{g_Q^*}

\section{Introduction}
Games played by a large number of people in a decentralized manner seem to be susceptible to 
malicious players and it becomes important to understand the performance degradation due to the presence of 
malice. A particular class of games which have come under focus are {\em congestion games}. An example where malicious activity can degrade the performance of a congestion network, is a Denial of Service attack on a web-server. In the past few years 
there have been studies [\cite{pom},\cite{byzantine},\cite{eqinmalusers}] investigating the effect of malicious players on the other rational but non-cooperative players of the game. 

We study the effect of malicious players in non-atomic congestion games from a centralized viewpoint. 
In particular, we investigate how adversely the presence of malicious entities can affect 
the {\em social optimum} of a congestion game. It is probably not very surprising that in general congestion games malicious
players can affect the social optimum badly (and we make it precise how adversely later). In the special case of linear load balancing games, however, we make the following
observations.

In a load-balancing game, we have $m$ parallel links joining a single pair of a source and a sink. One unit of flow is to be distributed on the links. Each link $i$ has a 
latency $l_i:[0,1]\to R_+$ which is assumed to be non-decreasing. If $z$ is the amount of flow through a link $i$, then the cost of sending an infinitesimal portion $\Delta z$ of the flow on that link is $l_i(z)\cdot\Delta z$. 
We consider two players: MAL and SOC controlling $\alpha$ and $(1-\alpha)$ fraction of the flow. 
The SOC player wishes to minimize the cost faced by his fraction of the flow, while 
the MAL player wishes to maximize this cost. We call this zero-sum game $G_{\alpha}$. 

Formally, a pure strategy of the SOC (and similarly, of the MAL) player is an allocation of flow on the $m$ links
given by the vector $\y = \{y_1,\cdots,y_m)$ (and similarly $\x = (x_1,\cdots,x_m)$) with $\sum_{k=1}^m y_k = (1-\alpha)$
(and similarly $\sum_{k=1}^m x_k = \alpha$). The cost faced by the SOC player on playing the strategy $\y$ while
the MAL player plays $\x$ is

$$C(\x,\y) = \sum_{k=1}^m y_kl_k(x_k+y_k)$$ 

The SOC player wishes to minimize this cost while the MAL player wishes to maximize it. This defines the game $G_\alpha$.\\

\textbf{Remark}: A mixed strategy $P$ is a probability distribution over pure strategies, \textit{i.e.} a player plays a mixed strategy $P$ is she plays a pure strategy $\z$ with probability $P(\z)$. Since we have a continuum of pure strategies in $G_\alpha$, one may think that an equilibrium in mixed strategies would imply an equilibrium in pure strategies by simply replacing a mixed strategy $P$ by a pure strategy $\hat{\mathbf{z}} = \sum_{\z}{P(\z)\z}$. Following example shows that this implication does not hold, at least by above trivial transformation. If SOC plays a mixed strategy $P$ and MAL plays $x$, then $$\mbox{Expected cost to SOC}\ =\ \sum_{\y} P(\y) \sum_{k=1}^m y_k l_k(y_k + x_k)\ =\ \sum_{k=1}^m \sum_{\y} ( P(\y) y_k l_k(y_k + x_k) )$$ 
Instead, if SOC plays $\hat{\mathbf{y}} = \sum_{\y}{P(\y)\y}$ in response to MAL's pure strategy $\x$, we get 
\begin{eqnarray}
\nonumber C(\x, \hat{\y})& = &\sum_{k=1}^m \hat{y}_k l_k(\hat{y}_k + x_k)\\
\nonumber & = & \sum_{k=1}^m \left(\sum_{\y} P(\y)y_k\right) l_k ((\sum_{\y} P(\y)y_k) + x_k)\\
\nonumber & = & \sum_{k=1}^m \left(\sum_{\y} P(\y)y_k\right) \sum_{\y} (P(\y)l_k(y_k + x_k))
\end{eqnarray}
Notice that the costs are different in the two cases. Therefore, our results for pure strategies are non-trivially stronger than their possible counterparts for mixed strategies - which themselves do not follow from classical theorems since the strategy space is a continuum.\\

Our main result is that there exists a pure strategy Nash equilibrium in a linear load-balancing game $G_\alpha$. 
Since the game is zero-sum, equivalently we have the following minimax theorem:
$$\max_\x ~\min_\y~ C(\x,\y) = \min_\y ~\max_\x ~C(\x,\y)$$
where $\x$, $\y$ are pure strategies of MAL and SOC respectively and $C(\x,\y)$ is the cost to SOC. 
This shows that the order of play does not matter and it makes sense to talk about optimal strategies for the players.

We feel this is bit surprising for many reasons. Firstly, since both players have an infinite number of strategies, it is not even clear why there should be an equilibrium
even in mixed strategies. Moreover, although our setting is that of a congestion game which have pure equilibria, the game $G_{\alpha}$ is not a congestion game itself. Finally, the linear latencies are required: there is an example of a two link game with simple convex latencies which has no pure strategy Nash equilibrium.

Another observation we make about $G_\alpha$ is that one of the optimal strategies for MAL  
is to route its $\alpha$-fraction of the flow {\em selfishly}, that is, what the equilibrium outcome would have been 
if this $\alpha$ fraction were controlled by infinitesimal non-atomic agents each trying to minimze its latency.
In other words, a malicious player can inflict \textit{no more harm} than a set of selfish agents. 
Possibly, the malicious player could use a self-destructive strategy to degrade the performance, but it somehow does not need to do so in linear congestion games. \\

Our next result bounds the {\em Cost of Malice} (CoM), a quantity which we use to measure the effect of malicious players on the system. Formally, CoM is defined as the ratio of the cost faced by the SOC player in an equilibrium of $G_\alpha$ to 
$(1-\alpha)\mbox{OPT}$, where $OPT$ is the socially optimum cost in the absence of malice, when the whole $1$ unit was 
controlled by the SOC player.
Thus CoM in a way measures the effect of the malicious players on the social optimum.
As probably expected, in general congestion games CoM can be unbounded. However, for linear load balancing games
we show it to be bounded by $(1+\alpha/2)$.

\subsection{Related Work}
In the computer science community, there have been a number of studies investigating the effect of malicious agents in congestion games.
The first work closest to our setting was by  Karakostas and Viglas \cite{eqinmalusers} who study how the presence of malicious
agents in a congestion game effects the price of anarchy. That is, how the ratio between the cost of agents in a Nash equilibrium and the social optimum, which the authors call the coordination ratio, change with the amount of malice in the system. 
More recently, Babaioff \textit{et al} \cite{pom} consider the malicious agent in the game along with the infinitesimal players 
and show that for general latency functions, the game need not have a pure Nash equilibrium, 
and they show the existence of a mixed Nash equilibrium.
Moreover, they define the {\em Price of Malice} as the rate of deterioration in 
the total performance of the remaining selfish players per unit flow of malice
and  show that there exist networks where the price of malice may indeed be negative, 
while on the other hand there are networks where the price could be quite large. 

As we note in the introduction, the focus of the works above were on the effect of malice on the {\em equilibrium} of the system: how the equilibrium degrades 
and how the game between the selfish agents and the malicious agent takes place. On the other hand we are more interested in how the presence of malice affects
the social optimum itself. Thus our work complements the works of Karakostas and Viglas \cite{eqinmalusers} and Babaioff \textit{et al} \cite{pom}. 

The study of malice has not been restricted to the congestion game setting. Moscibroda \textit{et al} \cite{byzantine} study the effect of malicious agents in a virus inoculation
game. Indeed, they also define a notion of price of malice as the ratio of the cost of an equilibrium setting with and without malicious agents. The definition
is not quite same as that of Babaioff \textit{et al} \cite{pom}. In fact, to avoid a third ``price of malice" definition, we call the effect of malice to the optimum, the ``cost of
malice'' instead.

Our work is in some sense similar to the work on Stackelberg 
strategies started by Roughgarden \cite{stackelberg} and later works by Karakostas and Kolliopoulos \cite{KK}, Swamy \cite{swamy} and Sharma and Williamson \cite{SW}.
In Stackelberg games there exists a leader
who controls some amount of flow which he plays first to which the remaining selfish players respond. It is instructive to compare this leader with our SOC player.
In fact, our proof of bounding the cost of malice goes via the strategy SCALE in the paper of Roughgarden \cite{stackelberg}.

\subsection{Notational Preliminaries}\label{sec:prelim}
We establish a few notations before going on. We use the function $br()$ to denote best-response.
Given MAL and SOC's plays $\x$ and $\y$, we write $\x\cong br(\y)$ if $\x$ is the best response to $\y$.
We denote SOC-MAL $:= \min_{\y} \max_{\x} C(\x,\y)$ as the minimum cost faced by the SOC player when
he plays first. We define MAL-SOC similarly. We will show the existence of a pure Nash equilibrium
by showing SOC-MAL = MAL-SOC. \\

We now review the properties of optimum and Nash flows in load balancing games. If an amount of 
flow $\beta \le 1$ is to be routed on the $m$-links and moreover this flow is controlled
by infinitesimal non-atomic agents, then the resulting (unique) equilibrium flow, $\hat{\x}$
has the following property:
$$\hat{x}_i > 0 \Rightarrow l_i(\hat{x}_i) \le l_j(\hat{x}_j)$$
We call this flow, $F_N(\beta)$. If a single player controlling the whole $\beta$ units of flow
plays $F_N(\beta)$, we say he is playing NASH. 

If the $\beta$ amount of flow is controlled by a single player wishing to minimize the total cost,
then the  optimum flow $\hat{\x}$ satisfies the following property:
\begin{equation}
\label{Equation5}
\hat{x}_i > 0\ \ \Rightarrow\ \ l_i(\hat{x_i}) + \hat{x}_il_i'(\hat{x}_i)\ \ \leq\ \ l_j(\hat{x_j}) + \hat{x}_jl_j'(\hat{x}_j)
\end{equation}
We denote the total cost faced by the player as $C_S(\beta)$.
Note that the cost of malice is defined as $CoM := \displaystyle{\sup_{(\x,\y): Nash(G_\alpha)}} \frac{C(\x,\y)}{(1-\alpha)C_S(1)}$

\section{Existence of Pure Nash equilibrium in Linear Congestion Games}
Suppose the latency of the link $i$ is given by $l_i(x) = a_ix+b_i$.
The following lemma characterizes the best response of the MAL player to a SOC player's play.

\begin{Lem}
\label{Lemma1}
Suppose the SOC player plays $\mathbf{y}$.
Then  $\mathbf{x}$ is MAL's best reponse to $\mathbf{y}$ if and only if
\begin{eqnarray}
\label{Equation1} x_i > 0 & \Rightarrow & a_iy_i \geq a_jy_j\ \ \ \ \ \forall j
\end{eqnarray}   
\end{Lem}

\begin{proof}
\noindent
{\em (Only if part)} For $x_i > 0$, we will prove that if $a_jy_j > a_iy_i$ for some $j$, then MAL can strictly increase the cost to SOC by moving his jobs from link $i$ to link $j$. Let $\mathbf{x'}$ be such that $x_k' = x_k$, for $k \neq i,j$, with $x_i' = 0$ and $x_j' = x_j + x_i$. We have,
\begin{eqnarray}
\nonumber 
C(\mathbf{x}', \mathbf{y}) - C(\mathbf{x}, \mathbf{y})  & =  & \sum_{k}{a_ky_k(x_k'-x_k)}\\
\nonumber & = & a_jy_jx_i - a_iy_ix_i \ >\ 0
\end{eqnarray}

\noindent
{\em (If part)} Let $L = \displaystyle\max_{k}{a_ky_k}$.\ For any $\mathbf{x}$ satisfying property \ref{Equation1}, we have $x_i > 0 \Rightarrow a_iy_i = L$. Therefore, $C(\mathbf{x}, \mathbf{y})\ =\  \sum_{k}{y_k(a_k(y_k + x_k) + b_k)} \ =\ \sum_{k}{y_kl_k(y_k)} + L\alpha$

That is, the cost to SOC when MAL responds with any $\mathbf{x}$ satisfying property \ref{Equation1} is the same, and has to be optimal
from the only-if part.  
\end{proof}

\begin{Thm}
\label{Theorem1}
In any linear load balancing game, 
there exists pure strategies $\x,\y$ of MAL and SOC such that $\x \cong br(\y)$ and $\y \cong br(\x)$.
Equivalently,  there exists a pure strategy Nash equilibrium in every load balancing game $G_\alpha$. 
\end{Thm}

\begin{proof}
Let $\mathbf{x} = F_n(\alpha)$, that is MAL's strategy is the NASH strategy. 
Then by definition of NASH (see Section \ref{sec:prelim}), we know that:
$x_i > 0\ \ \Rightarrow \ \ a_ix_i + b_i\ \leq\ a_jx_j + b_j,\ \forall j$

Let us now characterize SOC's best response: $\mathbf{y} \cong br(\mathbf{x})$. 
Note that $\mathbf{y}$ is the strategy OPT  with the induced latencies 
$\tilde{l}_i(y_i)\ =\ l_i(x_i + y_i)$. Therefore, 
\begin{eqnarray}
\nonumber 
y_i > 0 & \Rightarrow & \tilde{l}_i(y_i) + y_i\tilde{l}_i'(y_i)\ \leq\ \tilde{l}_j(y_j) + y_j\tilde{l}_j'(y_j)\ \ \forall j\\
\nonumber & \Rightarrow & 2a_iy_i + a_ix_i + b_i\ \leq\ 2a_jy_j + a_jx_j + b_j\ \ \forall j
\end{eqnarray}

We now claim that $\x$ is in fact a best response to any such $\y$. To show this, from Lemma\eqref{Lemma1} 
it is enough to show
\begin{eqnarray}
\nonumber x_i > 0 & \Rightarrow & a_iy_i \geq a_jy_j \ \ \ \ \forall j 
\end{eqnarray}

Consider $i$ such that $x_i > 0$ and any other $j$. If $y_j = 0$ then the above inequality is trivially true. 
Otherwise from the optimality property of $\y$, we have 
$2a_jy_j + a_jx_j + b_j \leq 2a_iy_i + a_ix_i + b_i$. 
But since $x_i > 0$, we have $a_ix_i + b_i \leq a_jx_j + b_j$. Therefore, $a_iy_i \geq a_jy_j$.
\end{proof}

Notice that one of the equilibrium pure strategies of the malicious player is NASH. Recall that NASH is equivalent to the equilibrium reached by a set of selfish agents, each controlling only an infinitesimal fraction of the flow. Hence, by the above min-max theorem, we conclude that MAL can do no more harm than a set of selfish agents.

\section{Bounding the Cost of Malice}
\label{Section3}
In this section, we bound the cost of malice for linear load balancing games.
Let \mbox{$\mathbf{s} = F_N(1)$} be the NASH flow of value 1 in a load-balancing game. 
By definition latencies are equalized by $\mathbf{s}$ on every link. Let $L$ be this common latency. 
Then the total cost of this flow,  $$C_N(1) = \sum_{k}{s_kl_k(s_k)} = L$$

\begin{Lem}
\label{malratparallel}
In linear load balancing games, MAL-SOC $\leq (1-\alpha) C_N(1)$
\end{Lem}
\begin{proof}
Suppose the player MAL plays \x. We demonstrate a strategy for the SOC player which gives him cost at most $(1-\alpha)C_N(1)$.
Define $S=\{i| \xi \geq \si\}$. Let \y\ be as follows 
$$\yi= \left\{ \begin{array}{ll}0 & \textrm{if $i \in S$}\\\thi & \textrm{otherwise} \end{array}\right.$$
where $\thi$ is such that $\thi + \xi \leq \si$. Since $\displaystyle\sum_{k\notin S}(s_i - x_i) \geq (1-\alpha)$, such a $\mathbf{y}$ exists. Moreover, since $\theta_i + x_i \leq s_i$ on each link used by SOC, the latency faced by SOC on any link is at most $L = C_N(1)$.
Therefore, $\mbox{MAL-SOC}\ \leq\ C(\x,\y)\ \leq\ (1-\alpha)\nashm$
\end{proof}

\noindent
Thus we get that no matter what \mal\ plays, \soc\ can always make sure that her cost is at most $(1-\alpha)C_N(1)$. 
Therefore, using the  theorem of Roughgarden and Tardos \cite{poa} which bounds $C_N(1)$ and $C_S(1)$,
in non-atomic linear load balancing games, we arrive at the following bound on the cost of malice.

\begin{Thm}
\label{43bound} In linear congestion games, SOC-MAL = MAL-SOC $\leq \frac{4}{3} (1-\alpha) \optm$.
Thus $\textrm{CoM} \le \frac{4}{3}$.
\end{Thm}

\noindent
Note that in the above theorem, the cost of malice {\em does not} depend on the amount of malice. In particular, 
even when there is no malice, we only get a bound of $4/3$. In what follows, we show that 
$\textrm{CoM} \le (1 + \alpha/2)$. This is a tighter bound when $\alpha \in [0, 2/3]$.

Consider the following  simple strategy for \soc\ player called \scale. This strategy was suggested by Roughgarden \cite{stackelberg}
in his study of Stackelberg scheduling strategies (although \scale~  was not a good strategy for his purposes). 
We are able to bound \soc's cost which in turn gives an upper bound on \socmal. 

\noindent
SOC is said to follow \scale\ if she simply plays $\y = (1-\alpha)\mathbf{y}^*$, where $\mathbf{y}^* = F_S(1)$, 
the socially optimum flow.  That is, she scales the strategy OPT down to a $(1-\alpha)$-flow. 
Abusing notation, we also denote by SCALE the value of $C(\x,\y)$ when SOC follows SCALE.

\begin{Thm} \label{thm:scale}
In linear load-balancing games, 
$$\scalem\ \leq\ (1 + \frac{\alpha}{2}) (1- \alpha) \optm$$
and therefore $CoM \le (1+\alpha/2)$.
\end{Thm}
\begin{proof}
By Lemma \ref{Lemma1}, we know that the best response of MAL is to pass all his flow through the link that maximizes $a_kx_k$. 
Without loss of generality, suppose \mal\ uses link $1$ to send all his flow. Then
$$\scalem = \sum_{k=1}^m y_k (a_ky_k + b_k) + \a1 y_1\alpha$$
Replacing $y_k$ by $(1-\alpha)y_k^*$ and using the fact that $y_k^*$ has cost $\opt$, we get
$$\scalem = (1-\alpha)^2 \optm + \alpha (1-\alpha) (\a1\yos + \sum_{k=1}^m b_ky_k^*). $$
Now we  prove $\a1\yos + \sum_{k=1}^m b_ky_k^* \leq \frac{3}{2}\optm$ which would complete the proof. 
As $\mathbf{y}^* = F_S(1)$, $\mathbf{y}^*$ obeys property \ref{Equation5}. For linear latencies, it translates into:
$$\yis > 0\ \ \Rightarrow\ \ 2\ai\yis +\bi\ \ =\ \ 2\a1\yos + \b1$$
Multiplying above equation by $\yis$, we get
$$\yis > 0\ \ \Rightarrow\ \ \ai{\yis}^2 +\frac{(\bi-\b1)}{2}\yis\ \ =\ \ \a1\yos\yis$$
Summing above equations over all $\yis > 0$ and invoking $\sum_{k=1}^m y_k^*= 1$, we get
$$\a1\yos = \sum_{k=1}^m a_k{y_k^*}^2 + \sum_{k=1}^m \frac{(b_k-\b1)}{2}y_k^*$$
Adding $\sum_{k=1}^m b_ky_k^*$ to both sides in above equation, we get
$$ \a1\yos + \sum_{k=1}^m b_ky_k^*\ \ =\ \ \optm + \sum_{k=1}^m \frac{(b_k-\b1)}{2}y_k^* \ \ \le\ \ \frac{3}{2}\optm$$
where the last inequality follows from the fact that $\sum_{k=1}^m (b_k - \b1)y_k^*\ \le\ \sum_{k=1}^m b_ky_k^* \le \optm$.
\end{proof}

\noindent
{\bf Remark:} We should state that the analysis of the SCALE strategy in Theorem \ref{thm:scale} is tight. Consider the following
two links with latencies $(1,Mx)$.  The social optimum is to send the flows in the ratio $\left(\frac{2M-1}{2M}, \frac{1}{2M}\right)$ and the optimum 
cost is $\frac{4M - 1}{4}$. The SCALE strategy would send flow as $\left(\frac{(2M-1)(1-\alpha)}{2M},\frac{(1-\alpha)}{2M}\right)$ and then MAL would send the $\alpha$ fraction
on the second link, leading to a cost $\frac{(1-\alpha)((4+2\alpha)M - 1)}{4M}$. Thus CoM = $\frac{(4+2\alpha)M - 1}{4M - 1}$ which goes to $(1 + \alpha/2)$ as $M$ goes to infinity.
On the other hand if SOC sends all its $(1-\alpha)$ flow on the first link, the cost of malice is $\frac{4M}{4M -1}$ which goes to $1$ as $M$ grows.

\section{Malice in General Congestion Games}
The zero-sum game between a malicious player MAL and a socially optimum player SOC is also well defined
for general congestion games as well. However, as we saw, 
the property of having an equilibrium in pure-strategies breaks down when we go to non-linear latencies
even in the load-balancing case. A question which arises is if the existence in pure strategies generalizes
to general linear congestion games. In particular, does this property hold in network congestion games.
We believe the answer to be negative although we do not have an example. Nevertheless, we do know of examples
where MAL's optimal strategy ceases to play selfishly. Rather, in most examples we consider, the optimal strategy 
of MAL is to route a flow which maximizes the minimum latency path. 

Moreover, in network congestion games, the cost of malice can be large. A simple example is a parallel link
$s$-$t$ network (which is the same as the load balancing game) with one extra path from $s$ to $t$ which uses
all the links. It is not hard to see that the best strategy of the MAL player is to route all its
flow on the extra path and then the cost-of-malice becomes proportional to the number of parallel links.
\bibliographystyle{alpha}
\bibliography{draft}
\end{document}